\documentclass[slac_one]{revtex4}
\usepackage{graphicx}
\usepackage{fancyhdr}
\newcommand{\sq}{{\vbox {\hrule height 0.6pt\hbox{\vrule width 0.6pt\hskip
3pt \vbox{\vskip 6pt}\hskip 3pt \vrule width 0.6pt}\hrule height 0.6pt}}}
\newcommand{\beq}{\begin{equation}}
\newcommand{\eeq}{\end{equation}}

\newcommand{\dslash}{\not{\hbox{\kern-2pt $\partial$}}}
\newcommand{\pslash}{\not{\hbox{\kern-2.3pt $p$}}}
 \newtoks\nslashfraction
 \nslashfraction={.13}
 \newcommand{\nslash}[1]{\setbox0\hbox{$ #1 $}
   \setbox0\hbox to \the\nslashfraction\wd0{\hss \box0}/\box0 }

\begin{document}

\title{Gravity and Strings}

%

\author{Steven B. Giddings}
\affiliation{Department of Physics, University of California, Santa Barbara, CA 93106}

\begin{abstract}

This is a broad-brush review of how string theory addresses several important questions of gravitational physics.  The problem of non-renormalizability is first reviewed, followed by introduction of string theory as an ultraviolet-finite theory of gravity.  String theory's successes also include 
predicting  both gauge theory and fermions.  The difficulty of extra dimensions becomes a possible virtue, when one notes that these lead to mechanisms to explain fermion generations, as well as a means to break the large gauge symmetries of string theory.  Finally, a long standing problem of string theory, that of fixing the size and shape of the extra dimensions, has recently been addressed and may shed light on the origin of the cosmological constant, the ultimate fate of our universe, as well as the question of why gravity {\it is} so weak.

\end{abstract}

\maketitle

\thispagestyle{fancy}


\section{PUZZLES OF GRAVITY}

In this lecture I plan to convey the basic ideas of string theory, particularly as they relate to some puzzles of gravitational physics.  At my home institution, string theory is a two to three quarter class, just to teach the foundation, so the best that can be done here is to give a very impressionistic view of some of the features of the theory, introducing some of the central ideas.  (For the same reason, I will only give a brief guide to the literature at the end, rather than inclusive references.) Nonetheless, I will also endeavor to bring the reader up to speed on some of the newest -- and most bizarre -- ideas of string theory, particularly pertaining to cosmology and the fate of the Universe.  We'll begin with the question of the day.

\begin{figure}
 \includegraphics[width=55mm]{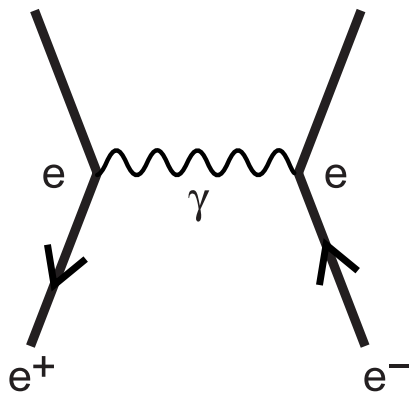}
 \caption{Single-photon exchange gives the leading contribution to Bhabha scattering.\label{fig1}}
 \end{figure}

\subsection{Why is gravity so weak?}

SLAC is famous for electron-positron scattering, and we know that a centrally important process here is Bhabha scattering, in which the final state is also an electron and positron, $e^+ e^- \rightarrow e^+ e^-$.  The leading contribution to this process is from one-photon exchange, fig.~1.  We know that the amplitude for this contribution is proportional to the square of the electron's charge, since there is an $e$ for each vertex where the photon attaches to the electron or positron:
\begin{equation}\label{bhabha}
{\cal A}_{e^+ e^-}^{\rm EM} \propto e^2 = \alpha\ .
\end{equation}
Here, and for the rest of the paper, we will use units so that $\hbar=c=1$.

 \begin{figure}
 \includegraphics[width=55mm]{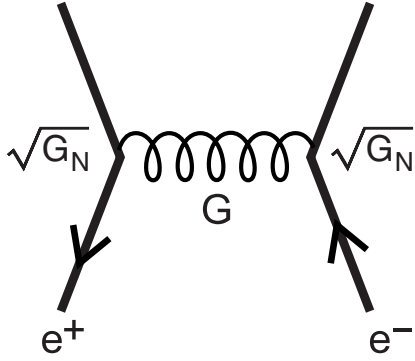}
 \caption{A subleading correction to the leading-order Bhabha scattering amplitude of  fig.~\ref{fig1} arises from one-graviton exchange.\label{fig2}}
 \end{figure}

However, when SLAC scatters electrons and positrons, there is a subleading contribution to Bhabha scattering arising from the process where the photon  is replaced by a graviton exchanged between the two particles, as in fig~\ref{fig2}.  This amplitude can be computed from the lagrangian for gravity,
\begin{equation}\label{EHlag}
S_{\rm grav} = \int d^4 x \sqrt{-g} \left[\frac{\cal R}{G_N} + {\cal L}(\psi_e,g)\right]\ .
\end{equation}
Here $g$ is the spacetime metric, $\cal R$ the curvature scalar, $G_N$ Newton's constant, and $\cal L$ the familiar Dirac lagrangian for the electron field $\psi_e$, in a general metric.  To get gravitational scattering from this, we expand the metric to exhibit fluctuations about flat space:
\begin{equation}\label{metexpan}
g_{\mu\nu}= \eta_{\mu\nu}+\sqrt{G_N} h_{\mu\nu}\ ,
\end{equation}
and the leading terms in a power series expansion of the action in $h$ take the form
\begin{equation}\label{actexp}
\sim\int d^4 x\left[ h^{\mu\nu} \sq_K h_{\mu\nu} + \sqrt{G_N} h^{\mu\nu} T_{\mu\nu} + {\cal O} (h^3)\right]\ 
\end{equation}
where $\sq_K$ is a generalized d'Alembertian for tensors, we've suppressed terms with higher powers of $h$ and/or more derivatives than two, and $T_{\mu\nu}$ is the stress tensor for the electron field.  

The leading interaction term in (\ref{actexp}) gives us one of the vertices in fig.~\ref{fig2}, and we see that it contains a factor of $\sqrt {G_N}$.  Dimensional analysis of (\ref{EHlag}) 
tells us that $G_N$ has mass dimension minus two, and, up to convention-dependent normalization, this defines the Planck mass scale,
\begin{equation}
G_N = M_{P}^{-2}\ .
\end{equation}
(In a theory with $n$ extra dimensions, we instead have $G_N = M_{P}^{-(2+n)}$.)  Given this fact, we see that whereas for the electromagnetic contribution to Bhabha scattering we had a dimensionless factor of $\alpha$, the amplitude for gravity contains in its place a factor proportional to $G_N$.  This factor must also be dimensionless, which means it must include a 
factor involving the characteristic energy scale $E$ of the  scattering process:
\begin{equation}\label{gravamp}
{\cal A}_{e^+ e^-}^{\rm grav} \propto G_N E^2= \left(E/M_P\right)^2\ .
\end{equation}
It's easy to see that the Planck mass is 
\begin{equation}
M_P \sim 10^{19} {\rm GeV}\ ,
\end{equation}
and so the gravitational amplitude is tiny even at TeV energies.  So today's question appears to morph into the question, why is the SLAC beam energy so low?

\subsection{The problem of predictivity}

\begin{figure}
 \includegraphics[width=85mm]{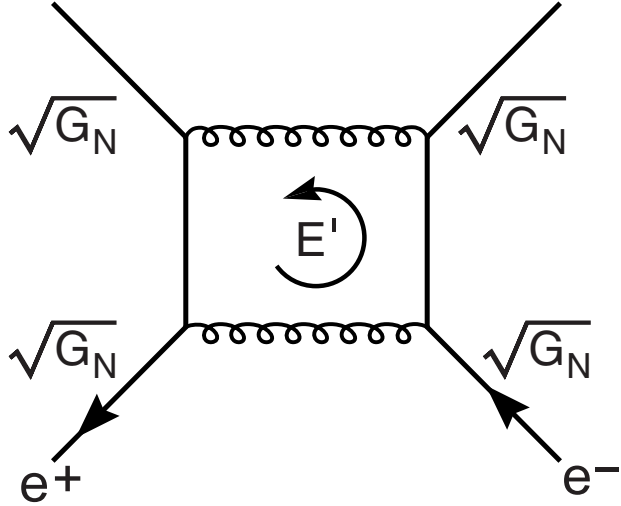}
 \caption{A gravitational loop diagram contributing to Bhabha scattering.\label{fig3}}
 \end{figure}

Once we are able to build a sufficiently high-energy accelerator, gravitational scattering will become important.  And, as in QED, Bhabha scattering will receive quantum corrections given by loop diagrams such as the one shown in fig.~\ref{fig3}.  Here we encounter quantum gravity's nasty surprise.  The one-loop amplitude shown has four interactions, hence two powers of $G_N$, and thus by dimensional analysis must have two more powers of energy.  Indeed, when the diagram is computed using the Feynman rules, we find that what enters is the loop energy $E'$, and  the correction to (\ref{gravamp}) behaves like
\begin{equation}
{\cal A}_{e^+ e^-}^{\rm grav,1} \propto G_N^2 E^2\int dE' E^{\prime}\ .
\end{equation}
This diverges badly at high loop energies/short distances.  Worse still, at higher-loop order, we have more powers of $G_N$, thus more powers of loop energy, and worse and worse divergences.  One could attempt to follow the usual program of renormalization, and absorb these divergences into the coupling constants of the theory.  But since there are an infinite number of divergences, present for any gravitational process, one needs an infinite number of coupling constants to renormalize the theory.   Practically, this means the theory is non-predictive:  in order to predict the outcome of various high-energy gravitational scattering experiments, we'd have to know the value of this infinite number of coupling constants, which would require infinitely many experiments to begin with.  Technically, we say the theory is  {\it non-renormalizable}. The same problem occurs whenever we have a theory with coupling constant with negative mass dimension. 

Physicists  have  encountered this problem previously, with the four-fermion weak interactions.  These interactions are described by terms in the lagrangian of the form
\begin{equation}\label{fourf}
{\cal L}_4 \sim G_F J_W^\mu J_{W\mu}
\end{equation}
where $J_W$ is a weak current, bilinear in fermions, and $G_F$ is Fermi's constant.  From dimensional analysis, we find that $G_F$ also has mass dimension minus two, and thus the four-fermi theory is as non-renormalizable and non-predictive as gravity.  But this is a problem we've seen resolved; we know that (\ref{fourf}) has the underlying structure
\begin{equation}\label{fourfa}
{\cal L}_4 \sim {g^2\over M_W^2}  J_W^\mu J_{W\mu}
\end{equation}
where $g$ is a {\it dimensionless} coupling constant, and $M_W$ is the mass of a heavy weak vector boson.  The lagrangian (\ref{fourfa}) is the low energy limit of an expression arising from exchange of the weak boson between two fermion lines, analogous to figs.~\ref{fig1},\ref{fig2}; $1/M_W^2$ arises as the low-energy limit of the propagator $\sim 1/(p^2 + M_W^2)$.  The underlying theory of spontaneously broken $SU(2)\times U(1)$ gauge symmetry is renormalizable, and thus predictive.

\subsection{Why do we care about predictivity in gravity?}

There are several reasons to be concerned about the breakdown of predictivity in quantum gravity.  The first is simply one of principle:  ultimately one can imagine building an accelerator that can scatter electrons at $E\geq M_P$, and then both single-graviton exchange as in fig.~\ref{fig2} and the higher loop corrections, like fig.~\ref{fig3} become important.  We should have a theory that describes this physics.  Part of the story is black hole formation, but we should say more.  This becomes even more important when we recognize that with large or warped extra dimensions, one might  encounter the fundamental Planck scale, and thus strong gravitational scattering, at the TeV scale.

A second motivation is that a complete theory of physics should encompass cosmology, and early in the history of the universe, typical particle energies approached or perhaps exceeded $M_P$.  Thus, to fully understand the initial conditions for our universe, we need a more complete theory.

Thirdly, there is exceptionally strong evidence that astrophysical black holes exist.  If we want to answer the question of what happens to an observer who falls into a black hole and reaches the high density/strong curvature region, we need to know more about the quantum mechanics of gravity.  Moreover, black holes evaporate, and small enough black holes would do so quite rapidly; to fully understand this process requires a predictive theory of quantum gravity.

Yet another possible motivation is that one place where we have attempted to combine basic quantum-mechanical notions and gravity has, so far, led to glaringly wrong predictions.  Quantum mechanics predicts a vacuum energy, which would contribute to the cosmological constant.  Any attempt to estimate the value, however, is off by tens of orders of magnitude from the value indicated by recent astrophysical observations, which is the same order of magnitude as the matter density in the universe.  We hope that a more complete understanding of gravity will help us with this problem

Finally, while the Standard Model is a useful theoretical guide to current experiment, it contains many parameters and certainly doesn't appear to be the final picture of physics.  Shorter-distance physics should provide boundary conditions that determine the parameters of the Standard Model, much as electroweak gauge theory determines the low-energy parameters of the four-fermion interaction.  Ultimately, specification of these boundary conditions will force us to understand gravity.

\section{LIGHTENING INTRODUCTION TO STRINGS}

\subsection{Gravity from strings}

\begin{figure}
 \includegraphics[width=150mm]{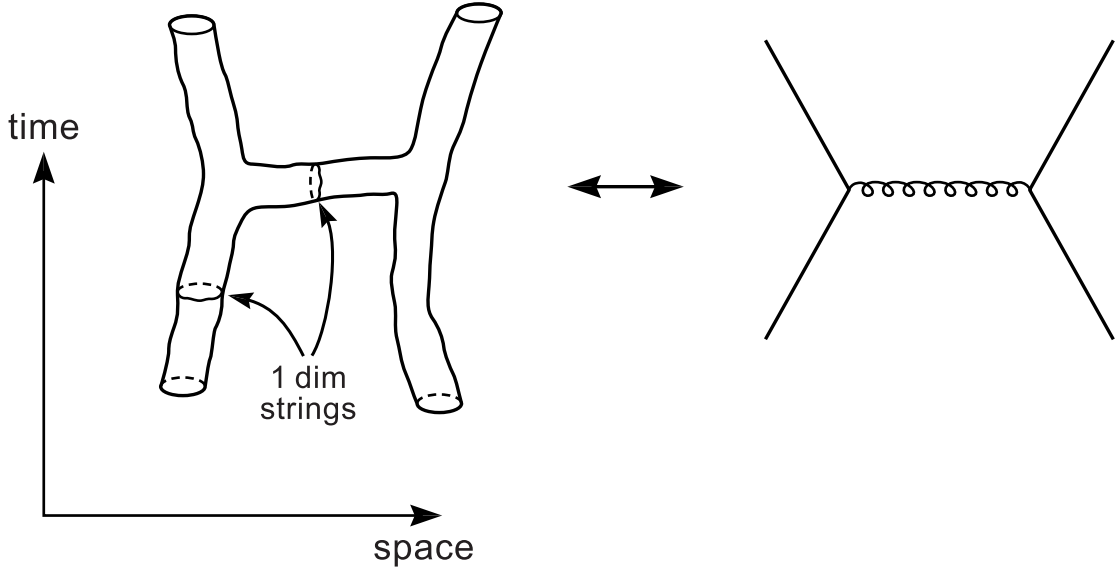}
 \caption{In string theory, the short-distance structure of the Feynman diagram of fig.~\ref{fig2} involves one-dimensional strings propagating through spacetime. \label{fig4}}
 \end{figure}

We've just outlined some of the most prominent reasons to seek a predictive theory of quantum-mechanical gravity.  Despite the importance of this problem, it is not fully solved.  Following the cue of the electroweak theory, the first place we might look is for a more fundamental field theory that gives an underlying renormalizable description of gravitational phenomena.  No such theory has been found.  But, through various historical accidents, a completely different approach to the problem has emerged.  This is a much more surprising modification of the theory at the {\it classical} level.  The basic picture is that the graviton exchange diagram of fig.~\ref{fig2} is derived from shorter-distance physics in which the basic objects are not particles, but rather {\it strings}.  We think of such a string as an infinitesimally thin filament of energy.  The incoming electron of the diagram, sufficiently magnified, is a small piece of string, and the graviton exchange corresponds to exchange of a loop of string, as shown in fig.~\ref{fig4}.

The starting point for a mathematical description of such a theory is that amplitudes, such as the one shown, are given by extending Feynman's sum-over-histories to a sum over string worldsheets interpolating between the string configurations in the initial and final states,
\begin{equation}\label{wsint}
{\cal A}_{\rm string\, string} \sim \int {\cal D}({\rm worldsheets}) e^{i{\rm Area}}\ .
\end{equation}
Here the quantity playing the role of the action is essentially the {\it Area} of the worldsheet swept out by the strings as they move through spacetime.

This prescription is simple but radical -- how do we know it reproduces gravity?

With more time we could compute the simplest amplitude of the form (\ref{wsint}), that for two-to-two scattering of the lowest excited state, call it  ``$T$," of oscillation of a string.  The result is
\begin{equation}\label{virshap}
{\cal A}_{TT\rightarrow TT} = \int {\cal D}({\rm worldsheets}) e^{i{\rm Area}} = { \Gamma\left(-{\alpha' s\over 4} -1\right) \Gamma\left(-{\alpha' t\over 4} -1\right) \Gamma\left(-{\alpha' u\over 4} -1\right)
\over \Gamma\left(-{\alpha' (s+t)\over 4} -2\right) \Gamma\left(-{\alpha' (t+u)\over 4} -2\right)
\Gamma\left(-{\alpha' (u+s)\over 4} -2\right)}\ .
\end{equation}
The ingredients of this amplitude, known as the ``Virasoro-Shapiro" amplitude, are as follows.  $\Gamma$ is the well-known generalized factorial.  The quantities $s$, $t$, and $u$ are the Mandelstam invariants, written in terms of the four incoming momenta $(p_1, p_2, p_3, p_4)$ as
\begin{equation}
s= -(p_1+p_2)^2\ ,\  t= -(p_1+p_3)^2\ ,\ u= -(p_1+p_4)^2\ ,
\end{equation}
and $\alpha'$ is a fundamental constant of the theory, with mass dimension minus two.  We can write
\begin{equation} 
\alpha' = M_S^{-2}
\end{equation}
where $M_S$ is known as the {\it string mass scale}.  

It's a straightforward exercise, using properties of the gamma function, to show that the amplitude (\ref{virshap}) has resonances, poles to be precise, at momenta such that
\beq\label{virshapres}
s\ ,\ t\ ,\ u = -{4\over \alpha'}\ ,\ 0\ ,\ {4\over \alpha'}\ , \ \ldots\ .
\eeq
Ignore the negative pole; it is eliminated in the supersymmetric version of the theory.  The pole at zero arises from a resonance corresponding to a {\it massless} state, and the higher poles to higher values of $m^2$.  

To infer the spin of the massless state we can take another limit of the amplitude (\ref{virshap}), namely $s\rightarrow \infty$ with $t$ fixed; this is the {\it Regge limit}.  In this limit, it is also a straightforward exercise to show
\beq
{\cal A}_{TT\rightarrow TT}\sim s^{2 + \alpha' t/2} f(t)\ ,
\eeq
where $f$ is a function just of $t$.  Basic resonance theory tells us that the exponent of $s$ corresponds to the spin $J$ of the intermediate state.  For a pole at $t=0$, we thus find $J=2$.  The physical interpretation of this is that the lowest excited state of vibration of the string has a quadrupole like waveform, characteristic of spin two, as can be confirmed from further analysis.  This state thus behaves just like a massless spin-two particle.

Now we can use a general result that goes back to Feynman:  any theory of an interacting spin two massless particle must describe {\it gravity}.  So string theory must reproduce gravitational physics.

This miraculous result has a couple of catches (which, we'll see, turn out to be bonuses).  First, the theory is only really sensible in $D=26$ spacetime dimensions, and otherwise has mathematical inconsistencies.  Second, it is really only the supersymmetric extension of what we've discussed that gives a well-defined theory; otherwise the first resonance in (\ref{virshapres}) is present and signals a tachyonic instability.  For the supersymmetric theory, the special dimension is instead $D=10$.

\begin{figure}
 \includegraphics[width=70mm]{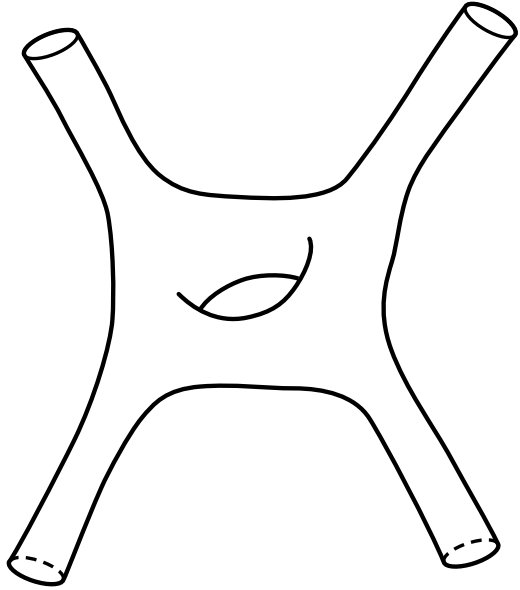}
 \caption{The string version of the one-loop diagram of fig.~\ref{fig3}.\label{fig5}}
 \end{figure}

With these caveats, we've made a major advance: the problem of non-renormalizability has apparently been cured!  For example the string version of the one-loop diagram of fig.~\ref{fig3} is shown in fig.~\ref{fig5}.  When computed, it's found that this diagram has no high-energy divergence; it is ultraviolet finite.  The reason for this is that, if we compute the diagram fig.~\ref{fig3} in a position-space representation, the singularity comes from coincident interaction points.  However, in the string diagram of fig.~\ref{fig5}, there are no special interaction points to give us a divergent coincidence.  At a deeper level, the ultraviolet divergence can be understood to be removed by certain {\it duality} symmetries, which relate potential ultraviolet divergences instead to infrared behavior.  This continues to higher loops, and a wonderful thing has occurred: string theory gives us a theory of gravity that is ultraviolet finite order-by-order in perturbation theory.

\subsection{The power of string theory}

String theory has offered us a solution to one major problem -- but that is only the beginning.  We know that non-gravitational forces are described by gauge theories, do these occur in string theory?

So far, we have focussed on closed strings.  But open strings, with endpoints, are also perfectly consistent -- with a little more structure.  It turns out that the mathematically consistent theories of open strings also require that a charge label be associated with each end of the string.  For illustration, the endpoints may transform in the $\bf N$ and ${\bar {\bf N}}$ of $SU(N)$.   

\begin{figure}
 \includegraphics[width=70mm]{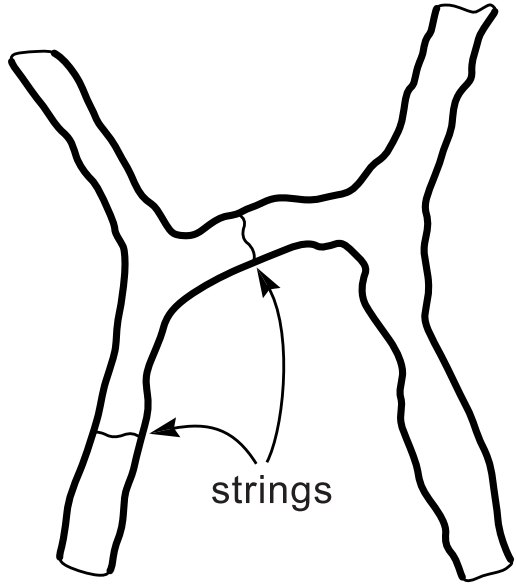}
 \caption{To compute open string scattering, one must sum over worldsheets with boundaries, corresponding to the worldlines of the endpoints of the strings.\label{fig6}}
 \end{figure}

Amplitudes with open strings are computed just like those for closed strings.  For example, two-to-two scattering is computed by summing worldsheet diagrams like those in fig.~\ref{fig6}.  Because the strings have endpoints, the worldsheet now has boundaries.  Once again the action is essentially just the area, and if we again denote the lowest state of the string as ``T," we find the amplitude
\beq
{\cal A}_{TT\rightarrow TT} = \int {\cal D}({\rm worldsheets}) e^{i{\rm Area}} = {\Gamma(-\alpha' s -1) \Gamma(-\alpha' t -1) \over \Gamma(-\alpha' s -\alpha' t-2) }\ ,
\eeq
the {\it Veneziano amplitude}.
Again the mass and spins of resonances in this amplitude follow from its poles and its Regge behavior, and in particular, aside from the tachyon that is eliminated in the supersymmetric version of the theory, the lowest excited state is zero mass and has spin one.  Moreover, taking into account the charges on the two ends of the string, we find states that behave just like non-abelian gauge bosons of the $SU(N)$ group.

At this point one might expect, like in field theory, that there are many possible theories with many different gauge groups.  Here is where part of the power of string theory enters -- string theory is a very tight mathematical structure, and it turns out that all but a handful of theories are {\it mathematically inconsistent}, suffering from quantum anomalies.  

The basic consistent string theories that we find are all supersymmetric and make sense only in ten spacetime dimensions.  One consists of both open and closed strings; in a theory of interacting open strings, two ends of a string can always join to give a closed string. The only consistent gauge group for this theory proves to be $SO(32)$.  The rest are closed string theories.  Two of these have no non-abelian gauge structure; they are called the type IIA and IIB theories, and differ by the chirality of the fermions that get introduced when incorporating supersymmetry.  Finally, there is yet another way to get non-abelian gauge groups, which is too complex to present here, but which yields the final two string theories, the heterotic string theories with gauge groups $E_8\times E_8$ and $SO(32)$.  That's it -- just five theories.

Another thing that the theory predicts is the existence of Dp branes.  These are extended $p$-dimensional objects.  So $p=1$ is a new kind of string, $p=2$ gives a membrane, and so on.  The ``D" stands for Dirichlet.  It turns out that open strings satisfy Dirichlet boundary conditions at a D brane, which physically means that open string ends get stuck on D branes.  The endpoints can move along the brane, but not transversely to it.  This actually gives another mechanism to get other gauge groups -- for example $SU(N)$ for strings moving along a stack of $N$ branes, but the corresponding gauge theories exist on the branes and not in the full nine spatial dimensions.  Na\"\i vely this suggests that there might be more string constructions, but at the same time, it turns out that the existence of branes help one to show that all five of the theories mentioned above are just different versions of the {\it same} underlying theory.  This ultimately is seen to happen through the existence of powerful duality symmetries which relate the theories.

Let's take stock so far.  The simple assumption that matter consists of strings, not particles, has produced gravity, and moreover this gravitational theory appears not to suffer the usual inconsistencies upon quantization.  Moreover, consistency of the theory requires supersymmetry, and hence fermionic matter, which is a first bonus of the theory. Finally, the theory naturally produces gauge symmetry.  It's quite amazing that most of the ingredients of known physics come out of one simple assumption.

But, at the same time, there are some apparent difficulties in describing the physical world.  The theory only makes sense in ten spacetime dimensions.  The gauge groups it produces are too big, and finally, while it predicts fermions, it is not clear how to get fermionic matter with the structure we see, for example generations.  

The remarkable thing is that, once we figure out how to solve one problem -- that of too many dimensions -- mechanisms that can solve the other problems {\it naturally appear}.

\section{HIDDEN DIMENSIONS}

\begin{figure}
 \includegraphics[width=130mm]{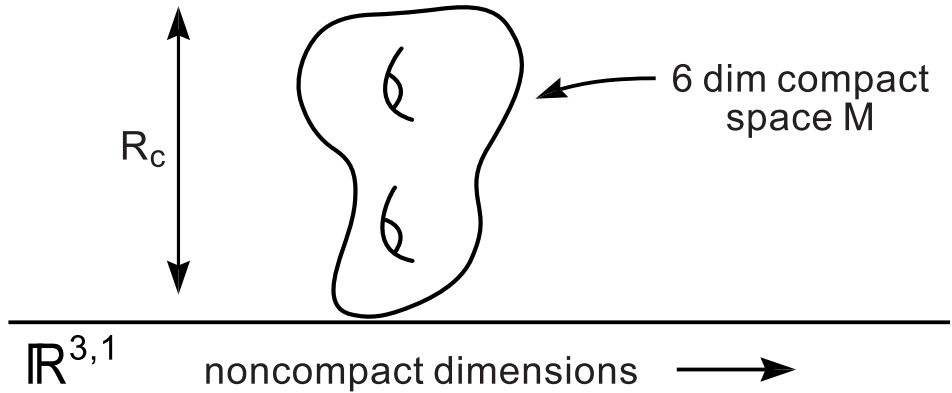}
 \caption{A representation of ten-dimensional spacetime:  four spacetime dimensions are extended and visible, and the remaining six are compactified on a small and hence invisible space, with size $\sim R_c$.\label{fig7}}
 \end{figure}

The basic idea in reconciling string theory with the four-dimensional reality of experience is that the ten dimensions may be configured so that six of them are folded into a small compact manifold $M$, and only the four that we see are extended over large distances (see fig.~\ref{fig7}).  If the characteristic size, $R_c$, of the manifold is assumed to be small compared to $1/{\rm TeV}$, then there is no reason that we would have unearthed this interesting structure.  (And, with the brane world idea, the extra dimensions could be even bigger.)

\begin{figure}
 \includegraphics[width=100mm]{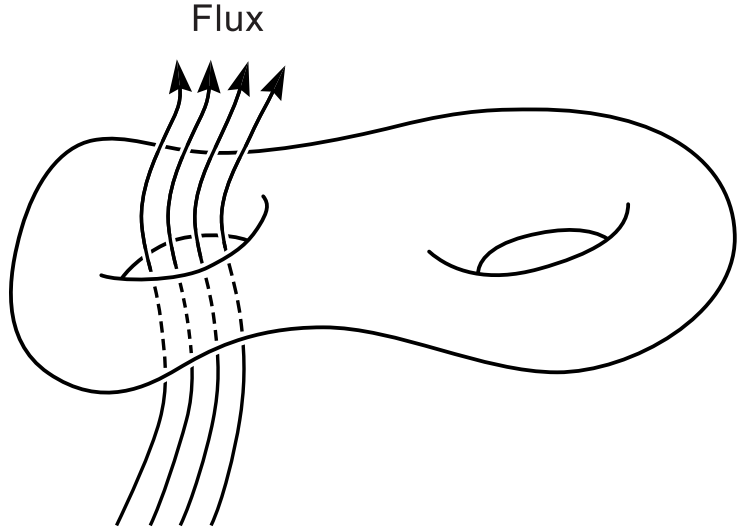}
 \caption{Flux lines trapped in  the topology of the compact manifold give a mechanism to break the large gauge groups of string theory.\label{fig8}}
 \end{figure}

Hiding the extra dimensions in this manner immediately yields a second bonus: a natural mechanism to break the large gauge groups we've encountered to something more realistic.  Specifically, the existence of non-trivial topology of the extra dimensions means that the gauge field configuration can include flux lines that are trapped in the topology, as illustrated in fig.~\ref{fig8}.  When present, these trapped flux lines, known as {\it Wilson lines}, break the gauge group.  One pattern of breaking looks like
\beq
E_8\rightarrow E_6 \rightarrow SO(10)\ ;
\eeq
$SO(10)$ is well-known to be a good group for grand unification, which can then break to the $SU(3)\times SU(2) \times U(1)$ of the Standard Model.  This is one possible mechanism to get the Standard Model; yet another way is from gauge symmetries on intersecting  D-branes.

A third bonus also can emerge from the presence of the extra dimensions: an answer to Rabi's old question, ``who ordered the muon?"  To see how extra dimensions can solve this problem of the generations, we realize that in the point-particle limit, a string configuration is described by a wavefunction $\psi(x,y)$ which is a function of the non-compact coordinates $x$ and compact coordinates $y$.  The wavefunction satisfies a generalized Dirac equation of the form
\beq\label{tendirac}
\nslash D_{10} \psi = \nslash D_4 \psi + \nslash D_6 \psi =0 \ ,
\eeq
 for the lowest oscillation state of the string.  Here the $\nslash D$'s are generalized Dirac operators with subscript indicating the dimension.  Then $\psi$ can be decomposed into normal modes $\psi_n$ of the compact operator, with eigenvalues $m_n$:
\beq\label{sixdirac}
\nslash D_6 \psi_n = m_n\psi_n\ .
\eeq
Eq.~(\ref{tendirac}) shows that the eigenvalue of the six-dimensional Dirac operator plays the role of the four-dimensional mass.  So, if eq.~(\ref{sixdirac}) has multiple eigenstates with the same charge, that leads to a replication of the spectrum of low-energy fermions, and could thus produce the generations
\beq
(\nu_e, e, u, d)\ ,\ (\nu_\mu, \mu ,c,s)\ ,\ (\nu_\tau, \tau, t, b)\ .
\eeq

So far, the story is quite remarkable.  We've assumed that matter is made of strings.  As output, we've found a quantum theory of gravity that is ultraviolet finite, gauge theories and thus the possibility to describe the standard model, fermions, a mechanism to produce generations, and, it turns out, scalars that can play the role of the Higgs.

Since it looks like all known physics can come out of string theory, it's been called a ``theory of everything," (TOE) although I prefer the phrase ``theory of all physics" (TOP).

\section{THE LANDSCAPE, AND OUR UNIVERSE'S FATE}

\subsection{Problems with moduli}

Before becoming too elated over all the successes of string theory, there's a critical question to ask:  
what fixes the compact space $M$?  Of course, this manifold must satisfy the equations of motion of string theory, which are, to leading approximation, the vacuum Einstein equations,
\beq\label{ricciflat}
R_{mn}(y) =0\ ,
\eeq
where $R_{mn}$ is the Ricci tensor.  The relevant manifolds are called {\it Calabi-Yau} manifolds.  

\begin{figure}
 \includegraphics[width=130mm]{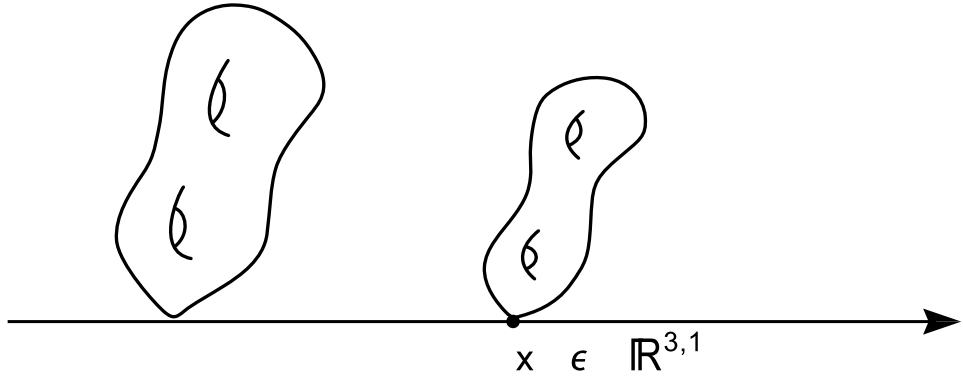}
 \caption{The size of the compact geometry can vary depending on location in our four spacetime dimensions.\label{fig9}}
 \end{figure}

Here we encounter a serious problem.  First, there are many topologies of Calabi-Yau manifolds, which represent discrete choices for the configuration of the extra six dimensions.  Moreover, there are many possible configurations of D branes wrapping the compact space.  But even worse, there are {\it continuous families} of Calabi-Yau manifolds, where the shape and size of the manifold varies continuously.  Moreover, these parameters may vary as a function of four-dimensional coordinate $x$.  
The simplest example is illustrated in fig.~\ref{fig9}, where the overall size of the manifold varies from point to point.   This variation is parametrized by a four-dimensional field $R(x)$ giving the characteristic size as a function of position.  Moreover, the fact that we have a solution of (\ref{ricciflat}) for any constant $R$ tells us that there is no potential for this field:  in the four-dimensional effective theory, it is a {\it massless} field.  One likewise finds other massless fields corresponding to various shape parameters of the manifold, for example the size of handles, etc.  These massless fields are all called {\it moduli} fields, and they are a disaster.  First, the lack of any prediction of the values of the moduli means that we lack predictivity:  parameters in the four-dimensional lagrangian, such as fermion masses and coupling constants will all vary with the moduli.  Worse still, the modulus fields interact with the other fields of the theory with gravitational strength.  Massless scalars with such interactions lead to fifth forces, time-dependent coupling constants, and/or extra light matter, none of which are seen experimentally.  

This represents a very serious problem for string theory, which has been present since the string revolution of 1984.  There have recently been some ideas about how to solve this problem, and relate it to another critical problem, that of the cosmological constant.  I'll summarize some of these ideas in the rest of the lecture.

\subsection{The landscape of string vacua}

We begin by reviewing one other ingredient of string theory:  $q$-form fluxes.  These are generalizations of electromagnetism, which has potential $A_\mu$, and anti-symmetric field strength
\beq
F_{\mu\nu}= \partial_\mu A_\nu - \partial_\nu A_\mu\ .
\eeq
Recall that the dynamics of electromagnetism is encoded in the Maxwell lagrangian, 
\beq 
S_{\rm Maxwell}\propto \int F^2\ .
\eeq
This structure can be generalized: consider a fully antisymmetric rank $q-1$ potential $A_{\mu_1\cdots\mu_{q-1}}$, and define an antisymmetric field strength
\beq
F_{\mu_1\cdots\mu_{q}} = \partial_{\mu_1} A_{\mu_2\cdots\mu_{q}} \pm {\rm permutations\ of}\ (\mu_1,\cdots,\mu_q)
\eeq
with action
\beq
S_q \propto \int F_q^2\ .
\eeq
It turns out that these $q$-form fields are present in string theory, and in fact, D-branes serve as sources for them much the same way an electron sources the electromagnetic field.

\begin{figure}
 \includegraphics[width=140mm]{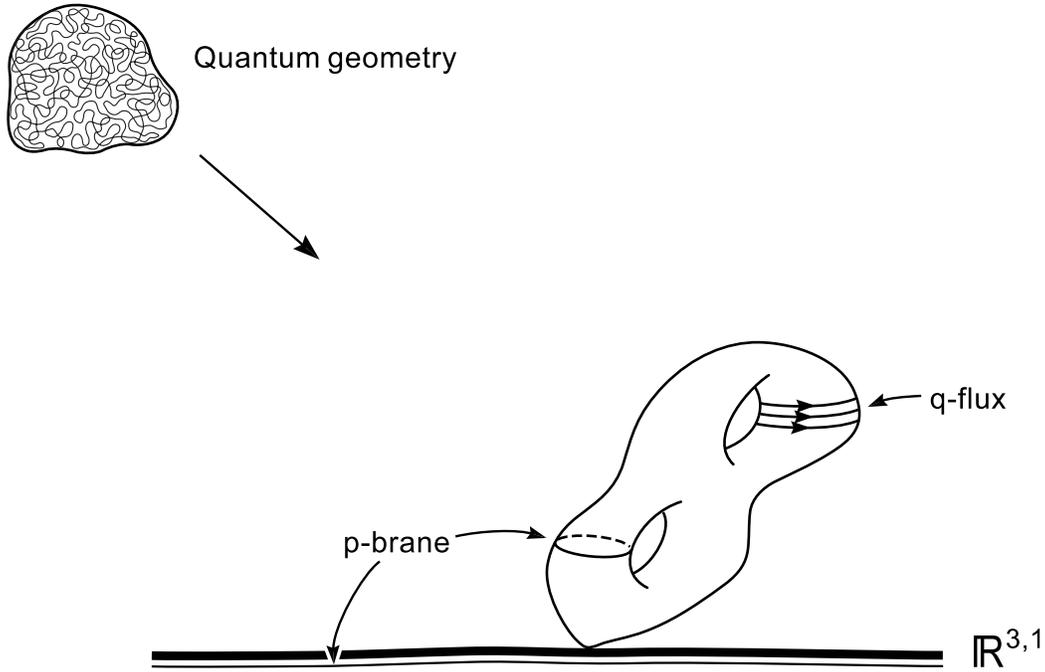}
 \caption{From an initial ``quantum geometry," as yet incompletely understood, our four dimensions and the compact dimensions should emerge.  In the process, branes and fluxes can be ``frozen" into the geometry.\label{fig10}}
 \end{figure}

Now, let us consider the Universe's evolution for its first few instants; a cartoon of this is shown in fig.~\ref{fig10}.  At the earliest times, we expect the very notion of classical spacetime geometry to break down, and be replaced by something more exotic.  As this evolves, we then might expect usual geometry to freeze out of this ``quantum geometry."  But, just like in any phase transition, remnants, such as defects, of the initial strongly fluctuating phase can be left behind.  For example, when it freezes out, the compact manifold could have some p-branes wrapped around some of its cycles.  Since we are interested in vacua that are approximately Poincar\'e invariant, we only consider the case where these branes are ``spacefilling," that is span the three spatial dimensions we see.  Likewise, the freeze-out of geometry can leave behind fluxes that are trapped in the six-dimensional topology.

Such trapped branes and fluxes then lead to an energy that depends on the shapes and sizes of the extra dimensions.  If we look just at the dependence on the overall scale $R$, a p-brane has energy that grows with the $p-3$-volume it wraps in the extra dimensions (since three of its directions are extended over visible dimensions),
\beq\label{braneE}
E_p\propto R^{p-3}\ .
\eeq
For fluxes, integrals of the form
\beq
\int F_q
\eeq
over q-submanifolds are fixed by quantization conditions, and so the energy behaves as
\beq\label{FluxE}
E_q=\int_M F_q^2 \propto {R^n\over R^{2q}}\ 
\eeq
where $R^n$ comes from the volume of the $n$ compact dimensions.
The  energies (\ref{braneE}), (\ref{FluxE}) then give an effective potential for $R$, in the theory used to summarize the physics seen by a four-dimensional observer.  
It turns out that a conversion factor is needed to express these energies in units used by a four-dimensional observer; when that's included, the effective potential is 
\beq
V(R) = \left[E_p(R) + E_q(R)\right]/R^{2n}\ ,
\eeq
with $n=6$ for the usual string case.
The resulting four-dimensional effective theory takes the form 
\beq
S=\int d^4 x\sqrt{-g} \left[ {\cal R} - k(\nabla \ln R)^2 - V(R) +\cdots \right]\ 
\eeq
where $k$ is a constant.

\begin{figure}
 \includegraphics[width=125mm]{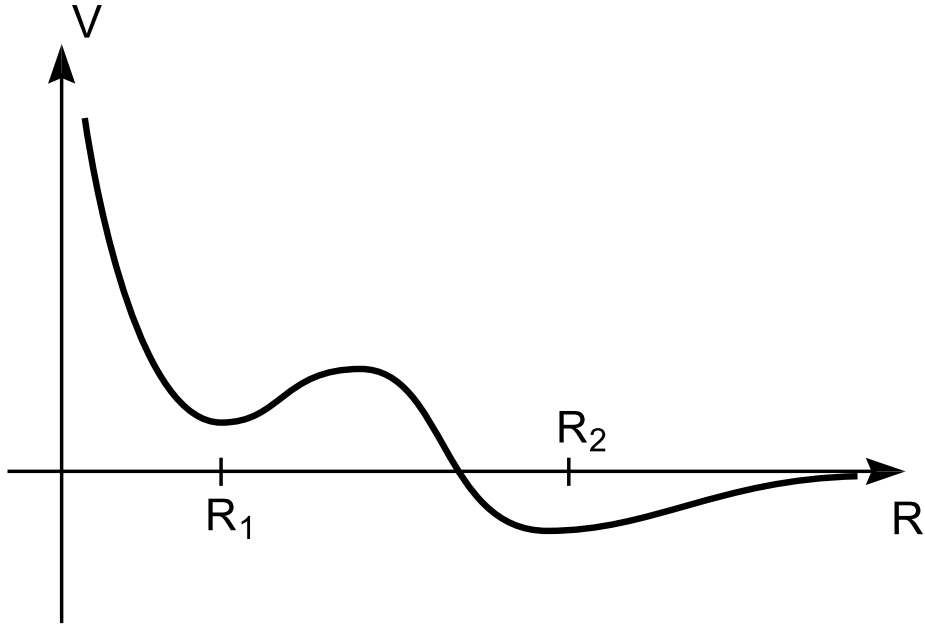}
 \caption{An example of a potential $V$ for a modulus  -- here the size $R$ of the compact dimensions.\label{fig11}}
 \end{figure}

The presence of such effective potentials implies that wrapped branes and fluxes, along with other more exotic effects, can therefore {\it fix} the moduli.  A rough sketch of an example of a potential is shown in fig.~\ref{fig11}.  This example has two minima.  The value of the potential at a minimum corresponds to a four-dimensional vacuum energy, that is {\it cosmological constant},
\beq
V(R_i) = \Lambda_{\rm eff}\ .
\eeq
So the negative minimum gives a negative cosmological constant, and the resulting vacuum cosmology is anti-de Sitter space.  Likewise, the minimum at positive potential gives a positive cosmological constant, which produces four-dimensional de Sitter space.

The consequences of this are very important: the energy from fluxes and branes trapped in the extra dimensions (and other related effects) may explain the presence of the dark energy, or cosmological constant. 

\begin{figure}
 \includegraphics[width=140mm]{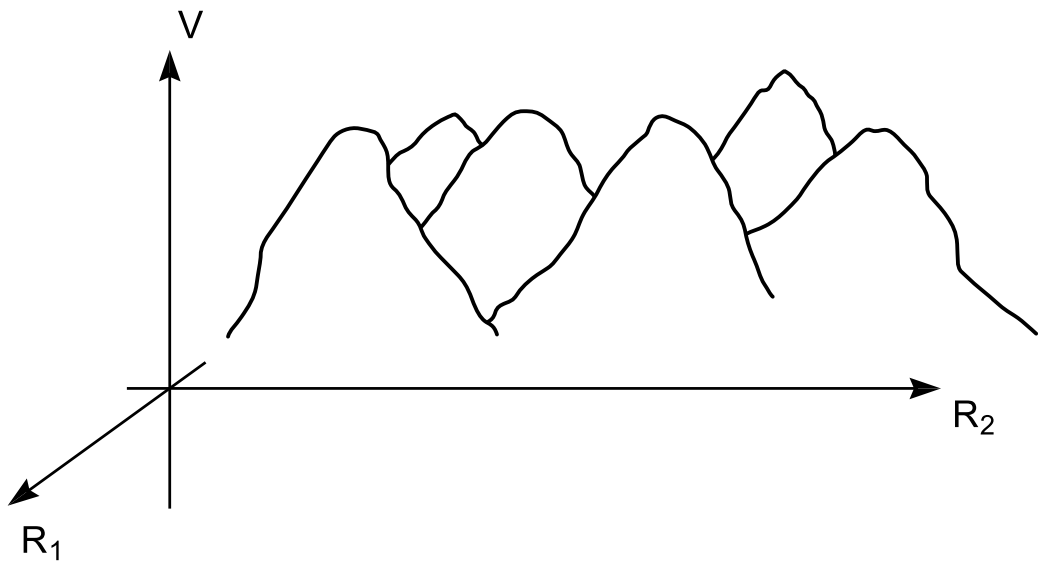}
 \caption{In general the space of string geometries is multi-dimensional, and fluxes, branes, and other related effects lead to a complicated  potential on this space.\label{fig12}}
 \end{figure}

The space of configurations of the extra dimensions is multi-dimensional, and in general there will be a complicated potential on this space, something like that sketched in fig.~\ref{fig12}.  This has been called the ``landscape" of string vacua.  Minima in this landscape correspond to (locally) stable four-dimensional vacua, with the potential at a minimum giving the cosmological constant.

One immediate problem in comparing this with observation is that, because the natural parameters entering the potential are set by string theory, typically the minima of the potential have values $V(R_i)\sim M_S^4\sim M_P^4$, which is about $10^{120}$ times the value $\Lambda_{\rm obs}$ that best fits recent astrophysical observations.

However, the space parametrizing the vacua is many dimensional, and the values of the minima of the potential are essentially randomly distributed.  So, if there are enough such vacua, which seems to be the case, then this random distribution of vacuum energies will yield {\it some} vacua with cosmological constant comparable to or smaller than $\Lambda_{\rm obs}$.

If one thinks about possible initial conditions for the Universe, it is quite plausible that they evolve into a state where the  landscape is populated so that different regions of the Universe are in different vacua in landscape.  Since $\Lambda_{\rm obs}$ is approximately the maximum allowed for galaxy formation, which would seem to be a necessary condition for life, we couldn't have evolved in a region of the Universe with a larger magnitude for the cosmological constant.  Since the distribution of cosmological constants is presumably dominated by the largest allowed value, this could serve as an explanation for the observed value $\Lambda_{\rm obs}$, arising from the {\it anthropic principle}.

This picture of string vacua is fairly new, and still being tested, but does seem quite plausible.  One potential issue is whether it emerges from a complete string theory analysis in a truly systematic fashion.  There are issues in carefully justifying the various approximations used in this analysis, and in the question of how to properly treat time-dependent solutions, such as cosmologies, in string theory, so work is still being done on this overall picture.

\begin{figure}
 \includegraphics[width=100mm]{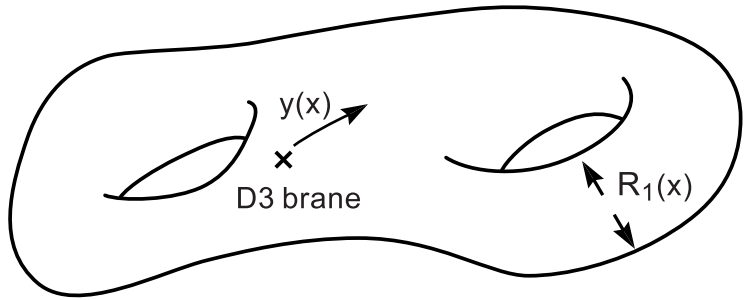}
 \caption{Motion of a brane on the internal space, or moduli fields, such as the size of a handle, can give candidates for the scalar field needed to drive inflation.\label{fig13}}
 \end{figure}

Assuming the landscape scenario survives, there are various other interesting aspects of it.  First, the basic picture gives us possible candidates for the fields necessary for inflation, namely the moduli fields and the fields describing motion of a D3 brane on the internal space.  (See fig.~\ref{fig13}.)  Moreover, brane collisions may have interesting effects, as in the ekpyrotic proposal.  Second, some points in the landscape may have large extra dimensions, or large {\it warping}, and could lead to scenarios of TeV scale gravity, where the true Planck scale is near the TeV energy scale.  In this case colliders like LHC might produce the ultimate exotica:  black holes.  Finally, anthropic ideas applied to the landscape have removed one fine-tuning, that of the cosmological constant.  This raises the question of the role of other fine tunings and hierarchies in nature, and their relationship to the cosmological constant.  For example, in the landscape it may be plausible to have the supersymmetry breaking scale much higher than the TeV scale, in which case superpartners may not be found anytime soon.  It may in fact be that anthropic considerations fix the small relative size of the Higgs mass as compared to the Planck mass.  If so, this ultimately answers the question we started with, ``why is gravity so weak?"  This is clearly a very interesting line of research, and 
debate continues on these and other important points.

\begin{figure}
 \includegraphics[width=130mm]{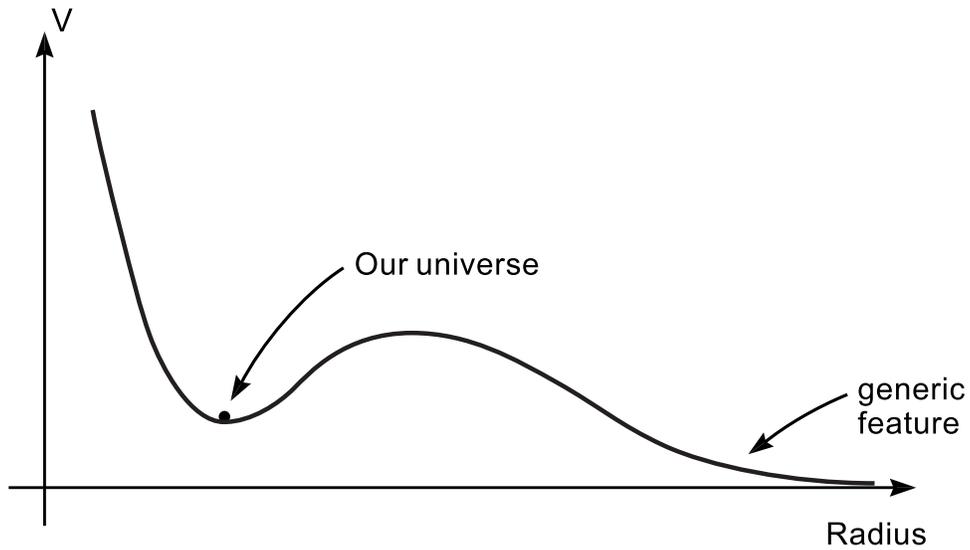}
 \caption{The potential for moduli must vanish for infinite volume of the compact dimensions, implying that a de Sitter region like our own is generically unstable to a decompactification transition.\label{fig14}}
 \end{figure}

A final fascinating point regards the final fate of our observable part of the Universe.  Since we observe a positive cosmological constant, we are apparently stuck at a positive minimum as shown in fig.~\ref{fig14}.  Now, it is possible to show on very general grounds that as the size of the extra dimensions goes to infinity, the potential always vanishes.  This feature is similar to the topographical transition from the Rockies to the great plains; the infinite plain tending to infinite volume  is a generic feature of the landscape.  This means that our region of the Universe is at best metastable, and will ultimately decay.  The generic decay is formation of a bubble of nine-dimensional space, which would then grow at the speed of light, consuming everything in its path.  We might refer to this process as {\it spontaneous decompactification} of the extra dimensions.  (Other more exotic, and equally deadly, decays of four-dimensions may also be present.)  Fortunately it will only happen on a timescale of $\exp\{10^{120}\}$  -- with a number this big, you can pick your favorite units.  But nonetheless, it's satisfying to have a possible understanding of the ultimate fate of our universe.

\section{SUMMARY}

Let's finish this lecture with a summary of some of the highlights:

\begin{itemize}

\item String theory apparently resolves the notorious non-renormalizability problem of quantum gravity.

\item It also predicts supersymmetry, and hence fermions; gauge theories; the possibility of generations of matter; and other features we observe in physics.

\item The possibility that string theory could incorporate all known physical phenomena means it is a candidate for a ``theory of all physics" (TOP).

\item  Nonetheless, it still faces some difficult problems, particularly that of how the size and shapes of the extra dimensions get fixed, known as the moduli problem.

\item One possible resolution of the moduli problem is the presence of branes or fluxes frozen into the geometry of the extra dimensions in their transition from an, as yet, mysterious ``quantum geometry."  

\item This mechanism for moduli fixing also predicts the presence of dark energy, specifically a cosmological constant.

\item It is likely that there are many possible string vacua lying in a multi-dimensional landscape; though details of this picture are being actively investigated.

\item The number of vacua of the landscape is so large that the cosmological constant may be fixed to its observed value by anthropic considerations.

\item Compactifications with branes and fluxes also give possible inflaton candidates, allow for the possibility of TeV scale gravity, and finally may make a very high supersymmetry breaking scale likely.

\item In this picture, a generic and ultimate instability of our four-dimensional universe is expansion of the extra dimensions in a process of spontaneous decompactification.

\end{itemize}

\section{GUIDE TO FURTHER READING}

Most of the material of sections 1-3 is contained in the basic texts, particularly 
\cite{Green:1987sp}\cite{Polchinski:1998rq}.  A broad class of string solutions that stabilize moduli using fluxes and branes was discovered in \cite{Giddings:2001yu}.  In many cases, these solutions stabilize all but one of the moduli, and then corrections to the leading-order action generate a potential for the remaining modulus.  This, together with the ingredient of adding an extra anti-D3 brane, was used in \cite{Kachru:2003aw} to find relatively explicit models of compactifications with all moduli stabilized, and moreover solutions with positive four-dimensional cosmological constant.  Building on this work, and on earlier work of Weinberg\cite{Weinberg:1987dv} and Bousso and Polchinski\cite{Bousso:2000xa}, Susskind\cite{Susskind:2003kw} coined the term ``landscape" and advocated the anthropic approach to fixing the cosmological constant in this framework.  There has been much subsequent literature on statistics of the vacua in the landscape and the possibility that anthropic considerations also allow a high supersymmetry breaking scale\cite{Arkani-Hamed:2004fb}; see \cite{Douglas:2004zg} for a review.  (A dissenting view on these developments is presented in \cite{Banks:2004xh}.) Finally, ref \cite{Giddings:2003zw} discusses the generic instability of our four-dimensional universe to spontaneous decompactification; more explicit discussion of this instability appears in \cite{Kachru:2003aw} and \cite{Giddings:2004vr}.

\begin{acknowledgments}
I'd like to thank the organizers of SSI for the opportunity to give this introductory lecture in such a well organized school.  This work was supported in part by Department of Energy under Contract DE-FG02-91ER40618, and
part of it was carried out during the Workshop on QCD and String
Theory at the Kavli Institute for Theoretical Physics, whose support is gratefully acknowledged.
\end{acknowledgments}


\end{document}